\documentstyle[aps,preprint]{revtex}

\renewcommand{\bar}[1]{\overline{#1}}
\begin{document}
\newcommand{\bra}[1]{\langle#1|}
\newcommand{\ket}[1]{|#1\rangle}
\newcommand{\be}[1]{\begin{equation}\label{eq:#1}}
\newcommand{\ee}{\end{equation}}
\newcommand{\req}[1]{(\ref{eq:#1})}
\newcommand{\bref}[6]{ #2, {\sl #3}, {\bf #4}, #5, (#6)}
\newcommand{\sigt}{\Sigma}
\draft
\tightenlines
\widetext
\title{
Absorption of radiation by small metallic particles: a general
self-consistent approach}
\author{M. Wilkinson$^1$ and B. Mehlig$^2$}
\address{\mbox{}$^1$
Department of Physics and Applied Physics,
John Anderson Building, 
University of Strathclyde,
Glasgow, G4 0NG, UK.\\
\mbox{}$^2$Theoretical Physics, University of Oxford,
1 Keble Road, Oxford, OX1 3NP,  UK}
\date{\today}
\maketitle{ }
\begin{abstract}
We introduce a theory for  
the absorption of electromagnetic radiation
by small metal particles, which generalises the
random phase approximation by incorporating 
both electric and magnetic dipole absorption within a
unified self-consistent scheme. 
We demonstrate the equivalence of the new approach to 
a superficially dissimilar perturbative approach. 
We show how to
obtain solutions to the self-consistent
equations using a classical approximation, 
taking into account the non-locality of the
polarisability and the conductivity tensor.
We discuss the nature of the self-consistent
solutions for diffusive and ballistic electron dynamics.
\end{abstract}
%\tableofcontents

\section{Introduction}   

The quantum theory of absorption of radiation by small metal
particles has proved to be a difficult area, partly because
it has not always been realised that the internal electric
field must be treated self-consistently, and partly because
of confusion about how to obtain the self-consistent field.
The literature is surveyed in [1-3], and the introductory
sections of [4] and [5] include a brief review of the
literature relevant to the present paper. This paper
extends a series of works [4-8] which have considered the
self-consistent potential from a semiclassical viewpoint.

First we clarify the regimes in which our results are
applicable. The following frequency scales are relevant to the 
interaction of small metal particles with radiation. The lowest
scale is given by the mean level spacing
$\Delta$, i.e., $\omega_\Delta = \Delta/\hbar$.
In $d$ dimensions $\omega_\Delta \propto a^{-d}$, where
$a$ is the characteristic size of the particle.
Another typical frequency scale is  given by
the inverse of the typical time taken for an electron to
traverse the particle. 
In systems with diffusive electron motion
it is given by $\omega_{\rm c} = D/a^2$, where $D$
is the diffusion constant.
In particles with ballistic electron motion,
$\omega_{\rm c}$ is given by the
inverse time of flight, $\omega_{\rm c} = v_{\rm F}/a$,
where $v_{\rm F}$ is the Fermi velocity. 
The highest frequency scales are the plasma frequency
$\omega_{\rm p}$ and the frequency $\omega_{\rm F}=E_{\rm F}/\hbar$
derived from the Fermi energy. 
In two dimensions, $\omega_{\rm p} \propto a^{-1/2}$, whereas
in three dimensions $\omega_{\rm p}$ is comparable to the 
Fermi frequency, which is independent of the size of the
particle. The frequency scales are therefore ordered
as follows
\be{1.1}
\omega_{\Delta} \ll \omega_{\rm c} \ll \omega_{\rm p} \le \omega_{\rm F}
\,.
\ee  
Our results are relevant to frequencies satisfying
$\omega \gg \omega_\Delta$ (which justifies the neglect
of quantum effects), and $\omega \ll \omega_{\rm p}$ (which
means that screening of external fields by polarisation
of the particle is significant).
There are six relevant length scales in the problem,
namely the wave length $\lambda$ of the external radiation, 
the linear dimension $a$ of
the particle, the skin depth
$\lambda_{\rm s}$, the Thomas-Fermi
screening length $\lambda_{\rm TF}$,
the Fermi wavelength $\lambda_{\rm F}$ 
and the mean free path $l$.
In the following it will be assumed that
\be{1.2} 
\lambda,\lambda_{\rm s} \gg a \gg  \lambda_{\rm TF} \sim \lambda_{\rm F}\,.
\ee
Both diffusive ($l\ll a$) and ballistic ($l \gg a$) 
dynamics will be discussed.

Under these conditions the absorption can be divided
into electric and magnetic dipole contributions.
In earlier papers these were discussed separately:
the electric dipole absorption was discussed for
various different situations in [4-7], and the magnetic
dipole absorption case was discussed in [8]. It is
desirable to have a unified description. In section II 
we develop a generalisation of the commonly used \lq random phase
approximation' [9] scheme which gives a uniform approach
to both electric and magnetic dipole absorption.
Most treatments of the electric dipole absorption 
coefficient are based upon calculating the imaginary 
part of the polarisability. References [4-7] used
an alternative approach, namely time dependent 
perturbation theory in the effective potential.
In section III we show that these apparently dissimilar 
approaches are equivalent.
 
We also discuss a unified approach to determining the
self-consistent field: we present some new results for
the ballistic case, where the bulk mean free path of the 
electrons is large compared to the dimensions of the 
particle. Section VI describes the form of the non-local 
polarisability, and section V discusses a general
semiclassical method for determining the self-consistent
field. We show that the self-consistent potential used
in [4,6], which treated electric dipole absorption ballistic
systems, was not correct for $\omega \gg \omega_{\rm c}$.
Section VI summarises the results, and presents an
argument indicating that the results of [4,6] are
nevertheless qualitatively correct.

\section{ Calculation of the absorption coefficient}

This section will discuss the general principles underlying
the calculation of the electromagnetic response.

\subsection{Formulation of the problem}

An electromagnetic wave induces currents which result in 
both electric and magnetic
polarisation of a conducting conducting particle. In what
follows we will only consider linear effects (where the 
polarisation is proportional to the applied field), and 
the externally applied field will be assumed to be 
uniform over the dimension of the particle.
We will only be concerned with the coefficients relating 
dipolar moments to the externally applied field: higher
moments will not be considered.
The electric dipole ${\bf d}$ and magnetic dipole ${\bf m}$ 
of a single particle are given by
\be{2.1}
$${\bf d}=\tilde \alpha\,{\bf E}_{\rm ext},\ \ \ \ 
{\bf m}=\tilde \beta\,{\bf B}_{\rm ext}
+\tilde \gamma \,{\bf E}_{\rm ext}
\ee
where ${\bf E}_{\rm ext}$ and ${\bf B}_{\rm ext}$ are the externally
applied electric and magnetic fields, $\tilde \alpha$
and $\tilde \beta$ are the electric and magnetic susceptibility 
tensors of the particle. The cross-susceptibility
$\tilde \gamma$ is not usually included. 
It is absent for spherical particles and
some other symmetric geometries, and when it is non-zero it
vanishes in the low-frequency limit.
We will ultimately give a completely general treatment,
showing that the cross-susceptibility makes no contribution
to the absorption, but for simplicity the cross-term will 
be dropped in the remainder of this introductory section.
The externally applied fields
are assumed to be multiplied by a factor of the form
$\exp(-{\rm i}\omega t)$, and the polarisability tensors
are understood to be functions of $\omega$ with complex valued 
components, because there may be a phase shift between the 
applied field and the response. For example, the actual value 
of the dipole moment at time $t$ is taken to be 
${\bf d}(t)={\rm Re}[{\bf d}\exp(-{\rm i}\omega t)]$.

These polarisations are detectable at a macroscopic level
in various ways: they alter the dielectric constant
and magnetic permeability of the medium in which
the particles are dispersed, and they may also
be detected by observing scattering and absorption
of radiation. The polarisability determines two 
processes which result in the 
attenuation of radiation, namely scattering and absorption of
energy. Both of these processes can be characterised conveniently
at the microscopic level by calculating the rate of loss of energy
from the incident beam due to interaction with a single particle: this
will be denoted by 
$\overline{dE/dt}$ [where $\overline{X}$ denotes the time
average of $X(t)$].
The two most commonly used measures
used to quantify these processes are the cross section per particle 
${\cal S}$ and the attenuation coefficient $\gamma $.
To relate the energy loss to the cross section, note that the 
energy density in an electromagnetic wave is 
${1\over 2}\epsilon_0{\bf E}^2$: the cross section is therefore
\be{2.2}
{\cal S}={2\over{\epsilon_0 c{\bf E}^2}}
\overline {dE\over{dt}}
\ee
where $\bar X$ denotes the time axerage of $X(t)$.
The attenuation coefficient $\gamma $ is defined by the 
expression $I=I_0\exp(-\gamma z)$, where $I$ is the intensity
at distance $z$ along the beam. The attenuation coefficient
is given by $\gamma ={\cal N}{\cal S}$, where ${\cal N}$
is the particle density.

At low frequencies the real part of the polarisability
approaches a constant (and the imaginary part approaches 
zero). It follows that at sufficiently low frequencies the
scattering cross section scales as $\omega^4$. It will be
shown that in the absorption cross section typically scales as
$\omega^2$, implying that absorption is expected to be the
dominant process at low frequencies.

The absorption of radiation can be related to the imaginary 
parts of the polarisability tensors: we will give a careful
explanation of this. Electron spin is not significant in this 
context, and the full Hamiltonian for the electrons is 
taken to be
\be{2.3} 
H=\sum_{i=1}^N {1\over{2m}}\bigl[ {\bf p}_i
-e{\bf A}_{\rm ext}( {\bf r}_i,t)\bigr]^2+V({\bf r}_i)
+\phi_{\rm ext}({\bf r}_i,t)
+{\textstyle{1\over 2}}\sum_{i=1}^N\sum_{{j=1}\atop{j\ne i}}^N
{e^2\over{4\pi \epsilon_0\vert{\bf r}_i-{\bf r}_j\vert}}
\ .
\ee
The externally applied electric and magnetic fields 
are considered to be spatially uniform, since the particle is 
small compared to the wavelength of the radiation, we ignore
the spatial dependence of the electric and magnetic fields,
and write:
\be{2.4}
\phi_{\rm ext}({\bf r},t)=e\,{\bf r}.{\bf E}_{\rm ext},\ \ \ 
\nabla \wedge {\bf A}_{\rm ext}({\bf r},t)={\bf B}_{\rm ext}(t)
\ .
\ee
In the case where the circularly symmetric gauge 
\be{2.5}
{\bf A}_{\rm ext}={\bf A}_{\rm rot}({\bf r},t)=
{\textstyle{1\over 2}}{\bf B}_{\rm ext}(t)\wedge {\bf r}
\ee
is used, the full Hamiltonian contains terms coupling the system to the 
electric and magnetic fields, of the following form:
$$\hat H(t)=\hat H_0+e\hat {\bf X}.{\bf E}_{\rm ext}(t)
+{e\over{2m}}\hat {\bf L}.{\bf B}_{\rm ext}(t)
+O({\bf B}_{\rm ext}^2)$$
\be{2.6}
\hat {\bf X}=\sum_{i=1}^N \hat {\bf r}_i,\ \ \ 
\hat {\bf L}=\sum_{i=1}^N {\bf r}_i\wedge {\bf p}_i
\ee
where $\hat {\bf X}$ and $\hat {\bf L}$ are the total 
dipole operator and total angular momentum operators, and 
$\hat H_0$ is independent of time.

To facilitate the calculations we will consider
ensemble averages of quantities: if the electron motion is ergodic, this
is the microcanonical average, and in general the ensemble
is defined by the region of phase space explored
by the dynamics. Angle brackets will be used for the 
appropriate ensemble average. 
For a general choice of gauge the instananeous rate of 
absorption is then
\be{2.7}
\biggl\langle{dE\over {dt}}\biggr\rangle
=\biggl\langle{\partial H\over {\partial t}}\biggr\rangle
=e\biggl\langle\sum_{i=1}^N{\bf v}_i(t).\bigl[
{\bf E}_{\rm ext}(t)+{\bf E}_{\rm ind}(t)\bigr]\biggr\rangle
=\int d{\bf r}\ {\bf j}({\bf r},t).
\bigl[{\bf E}_{\rm ext}(t)+{\bf E}_{\rm ind}({\bf r},t)\bigr]
\ee
where ${\bf v}_i$ is the velocity of the $i^{\rm th}$ electron,
${\bf E}_{\rm ind}={\rm i}\omega {\bf A}_{\rm ext}$
is the electric field induced by the varying magnetic field, and
\be{2.8}
{\bf j}({\bf r},t)=e\biggl\langle \sum_{i=1}^N 
{\bf v}_i(t)\,\delta [{\bf r}-{\bf r}_i(t)]\biggr\rangle
\ee
is the current density within the particle.
In the special case where the circularly symmetric gauge
is used this reduces to:
\be{2.9}
\biggl\langle {dE\over {dt}}\biggr\rangle=
\Biggl\langle{\partial H\over {\partial t}}\Biggr\rangle
=e\langle \hat {\bf X}\rangle .\dot {\bf E}_{\rm ext}(t)
+{e\over{2m}}\langle \hat {\bf L}\rangle .\dot {\bf B}_{\rm ext}(t)
\ee
where $\langle \hat {\bf X}\rangle$ and 
$\langle \hat {\bf L}\rangle$ are suitable averages
of the centre of mass and angular momentum operators. 
It is impractical to calculate these averages from 
the full Hamiltonian \req{2.3}, and in the next sub-section
it will be shown how they may be estimated using 
an {\sl effective} Hamiltonian, containing effective fields 
${\bf A}_{\rm eff}$ and $\phi_{\rm eff}$. At this stage
we will only assume that these averages are proportional
to the applied fields. These quantities 
$\langle \hat {\bf X}\rangle$ and $\langle \hat {\bf L}\rangle$
are related to the electric and magnetic dipole moments
${\bf d}$ and ${\bf m}$:
\be{2.10}
{\bf d}=e\langle {\bf X}\rangle, \ \ \ 
{\bf m}={e\over m}\langle {\bf L}\rangle 
\ .
\ee
The rate of absorption is
obtained by substituting for the time dependence of a monochromatic
field using \req{2.1}, and ignoring the cross term:
\begin{eqnarray}
\biggl\langle {dE\over{dt}}\biggr\rangle
&=&-{\rm Re}[{\bf d}\exp(-{\rm i}\omega t)]\,
{\rm Re}[{\rm i}\omega\, {\bf E}_{\rm ext}\exp(-{\rm i}\omega t)]
\nonumber\\
&&\hspace*{2cm}-
{\textstyle{1\over 2}}
{\rm Re}[{\bf m}\exp(-{\rm i}\omega t)]\,
{\rm Re}[{\rm i}\omega \,{\bf B}_{\rm ext}\exp(-{\rm i}\omega t)] \ .
\label{eq:2.11}
\end{eqnarray}
Averaging over time gives the general form for the rate of absorption
\be{2.12}
\biggl\langle \overline{dE\over{dt}}\biggr\rangle
={\textstyle {1\over 4}}\omega
{\bf E}_{\rm ext}^+(\tilde\alpha-\tilde\alpha^+) {\bf E}_{\rm ext}
+{\textstyle{1\over 8}}\omega 
{\bf B}_{\rm ext}^+(\tilde\beta-\tilde\beta^+) {\bf B}_{\rm ext} 
\ .
\ee
In the case where the polarisability tensor is isotropic, and 
the radiation field is plane polarised, this 
expression becomes
\be{2.13}
\biggl\langle\overline{{dE}\over{dt}}\biggr\rangle=
{\textstyle{1\over 2}}\omega \,{\rm Im}
\bigl[\alpha_{ii}(\omega)\bigr]\,\vert{\bf E}_{\rm ext}\vert ^2
+{\textstyle{1\over 4}}\omega \,
{\rm Im}\bigl[\beta_{ii}(\omega)\bigr]\,\vert {\bf B}_{\rm ext}\vert^2
\ee
We note that under the assumptions listed above, the absorption
is expressed as the sum of two terms, which are naturally referred
to as the electric and magnetic dipole absorption. Our final
result will not neglect the magnetic dipole moment which 
may be induced by the electric field, but we will show that
within the framework of our self-consistent approximation
scheme the cross term in \req{2.1} makes no contribution to the 
absorption. The energy absorbed does not accumulate in the 
system electronic system: most of it is eventually transformed 
into heat by interaction with phonons.

\subsection{Self-consistent fields}

The Hamiltonian will be approximated by an effective Hamiltonian, 
in which the electrons move independently.
The direct interaction with the magnetic
field via electron spin can also be neglected, and the 
effective Hamiltonian is of the form
\be{2.14}
\hat H_{\rm eff}=\sum_{i=1}^N{1\over{2m}}
[\hat {\bf p}_i-e{\bf A}_{\rm eff}({\bf r}_i,t)]^2
+V_{\rm eff}({\bf r}_i)+\phi_{\rm eff}({\bf r}_i,t)
\ .
\ee
The response of the system is determined by the 
interaction of the electrons with the electric field
inside the particle, which is described by the effective
potentials ${\bf A}_{\rm eff}$ and $\phi_{\rm eff}$.
The effective potentials are themselves determined
by the distribution of charge within the particle.
The external
magnetic field is also augmented by an induced magnetic field
which is produced by the action of the currents which flow in
order to establish the electric polarisation. Provided the
particle is sufficiently small, the induced magnetic field
can be neglected, and our self-consistent theory will
yield and equation for the electric field ${\bf E}({\bf r},\omega)$
within the particle, which is related to the time-dependent 
electric field as follows:
\be{2.15}
{\bf E}({\bf r},\omega) = \int_{-\infty}^{\infty}
\!dt\, {\bf E}({\bf r},t)\, \exp(-{\rm i}\omega t)
\ .
\ee
This field satisfies the Maxwell equations 
\be{2.16}
\nabla .{\bf E}={\rho \over{\epsilon_0}}\ ,\ \ \ 
\nabla \wedge {\bf E}={\rm i}\omega {\bf B}_{\rm ext}\,.
\ee
The electric field produced by induction when the external 
electric field is zero will be denoted ${\bf E}_{\rm ind}$.
The total effective electric field is
\be{2.17}
{\bf E}({\bf r},\omega) = {\bf E}_{\rm ext}({\bf r},\omega)
+ {\bf E}_{\rm ind}({\bf r},\omega)
+% e^{-1} 
\nabla \phi_{\rm pol}({\bf r},\omega)
\ .
\ee
The uniform external electric field
satisfies $\nabla.\,{\bf E}_{\rm ext}=0$, and can be derived 
from an external potential: 
\be{2.18}
{\bf E}_{\rm ext}({\bf r},\omega) 
=% e^{-1}
\nabla\phi_{\rm ext}({\bf r},\omega ) 
\ .
\ee
The potential $\phi_{\rm pol}$ results from polarisation
of the particle due to the external electric field,
and is given by 
\be{2.19}
\phi_{\rm pol}({\bf r},\omega)=%e^{-1} 
{1\over{4\pi \epsilon_0}}
\int d{\bf r}'\ {\rho_{\rm pol}({\bf r}',\omega)
\over{\vert {\bf r}-{\bf r}'\vert}}
\ee
where $\rho_{\rm pol}({\bf r},\omega)$ is the charge density
resulting from polarisation induced by the external electric
field, but excluding any polarisation which may result from the
induction field.
It will be convenient to express \req{2.17}
using the notation
\be{2.20}
|{\bf E}) = %e^{-1}
\nabla|\phi_{\rm ext}) + |{\bf E}_{\rm ind})
+ %e^{-1} 
\nabla \widehat U |\rho_{\rm pol})
\ee
where $\widehat U$ is an operator defined by \req{2.19}, 
acting on the \lq field vector' $|\rho_{\rm pol})$. 
The dependence upon frequency
will usually be shown explicitly for operators, but not
for field vectors.  

The current density ${\bf j}({\bf r},\omega)$ flowing in the sample to
build up the charge density $\rho({\bf r},\omega)$
may be assumed to be linearly related
to the electric field ${\bf E}({\bf r},\omega)$ in the sample
\be{2.21}
{\bf j}({\bf r},\omega)=\int d{\bf r}'\ 
\sigt ({\bf r},{\bf r}';\omega)\,
{\bf E}({\bf r}',\omega)\,.
\ee
In condensed notation, we write
\be{2.22}
\vert {\bf j}) = \widehat \sigt(\omega)\, \vert {\bf E})
\ee
where $\widehat \sigt(\omega)$ is the conductivity
operator. The non-local conductivity tensor 
$\sigt({\bf r},{\bf r}',\omega)$ is related to the 
non-local polarisability operator $\Pi ({\bf r},{\bf r}',\omega)$, which
gives the charge density induced by a potential
$\phi ({\bf r})$: we write
\be{2.23}
\rho({\bf r},\omega)=\int d{\bf r}'\ \Pi ({\bf r},{\bf r}';\omega)
\,\phi ({\bf r}',\omega)
\ee
or in condensed notation
\be{2.24}
\vert \rho )=\widehat \Pi (\omega)\, \vert \phi )
\ .
\ee
The polarisability operator $\widehat \Pi(\omega)$ 
can be related to a non-local
conductivity tensor $\widehat\sigt(\omega)$ 
by a continuity relation. 
Using \req{2.21}  and applying
the continuity equation, we find (with summation over repeated
indices implied)
\be{2.25}
0=\int d{\bf r}'\ \biggl[
{\rm i}\omega\,\Pi({\bf r},{\bf r}';\omega)\phi ({\bf r}')
-\nabla_i\Sigma_{ij}({\bf r},{\bf r}';\omega)
\nabla_j'\phi ({\bf r}')\biggr]
\ee
assuming that the normal component of $\widehat\sigt(\omega)$ 
vanishes on the boundary.
Upon integration by parts, after noting that the resulting
equation is valid for any field $\phi ({\bf r},\omega)$, we find
\be{2.26}
{\rm i}\omega\,\widehat\Pi(\omega) = 
-\overrightarrow{\strut\nabla} \widehat\sigt(\omega) 
\overleftarrow{\strut\nabla}
\ .
\ee
The quantities $\widehat \Pi(\omega)$ and $\widehat \sigt(\omega)$
enable \req{2.20} to be expressed in terms of the electric field alone,
yielding a self-consistent equation. We write
\be{2.27}
|{\bf E}) = %e^{-1} 
\nabla |\phi_{\rm eff}) + |{\bf E}_{\rm ind})	
\ee
where $\phi_{\rm eff}({\bf r},\omega)$ is an 
effective potential. We consider the solutions for the
field ${\bf E}_{\rm ind}$ and the potential $\phi_{\rm eff}$
separately. The charge induced by the field ${\bf E}_{\rm ind}$
is
\be{2.28}
\vert \rho_{\rm ind})={1\over {{\rm i}\omega}}
\nabla \widehat\sigt (\omega).\, \vert{\bf E}_{\rm ind})
\ .
\ee
Applying the first of the Maxwell equations \req{2.16} and using \req{2.28} 
gives
\be{2.29}
\nabla .\bigl[\vert {\bf E}_{\rm ind})
-{\rm i}\omega \widehat \sigt(\omega)\vert {\bf E}_{\rm ind})\bigr]=0
\ee
which is the self-consistent equation which must be solved
for the field ${\bf E}_{\rm ind}$.
For the effective potential, we find
\be{2.30}
\vert \phi_{\rm eff})
=\vert \phi_{\rm ext})+\widehat U\widehat \Pi (\omega)
\vert \phi_{\rm eff})
\ .
\ee
This self-consistent equation is sometimes referred to as the
\lq random phase approximation' [9]. 
Equations \req{2.29} and \req{2.30} must must
be solved for the self-consistent fields. We will consider 
semiclassical methods for solving them in section V.

\subsection{The rate of energy absorption}

The rate of energy absorption is given by \req{2.7}. Averaging
over time gives
\be{2.31}
\biggl\langle \overline{{dE\over{dt}}}\biggr\rangle
={\textstyle{1\over 2}}\, {\rm Re} 
\int\!d{\bf r}\ {\bf j}^\ast({\bf r},\omega).
{\bf E}_{\rm ext}+
{\textstyle{1\over 2}}\, {\rm Re} 
\int\!d{\bf r}\ {\bf j}^\ast({\bf r},\omega).
{\bf E}_{\rm ind}({\bf r},\omega)
\ .
\ee
In condensed notation this will be written, by analogy with Dirac
notation, as
\be{2.32}
\biggl\langle \overline{{dE\over{dt}}}\biggr\rangle
={\textstyle{1\over 2}}\, {\rm Re}\, ({\bf j}|{\bf E}_{\rm ext})
+{\textstyle{1\over 2}}\, {\rm Re}\, ({\bf j}|{\bf E}_{\rm ind})
\ .
\ee
Using \req{2.20},
\begin{eqnarray}
\biggl\langle \overline{{dE\over{dt}}}\biggr\rangle
&=&{\textstyle{1\over 2}}\, {\rm Re}\, ({\bf j}\vert {\bf E})
-{\textstyle{1\over 2}}\, {\rm Re}\, ({\bf j}\vert 
\nabla \widehat U\vert \rho_{\rm pol})
% \nonumber\\
={\textstyle{1\over 2}}\, {\rm Re}\, ({\bf j}\vert {\bf E})
-{\textstyle{1\over 2}}\, {\rm Re}\, {\rm i}\omega 
(\rho_{\rm pol}\vert \widehat U \vert \rho_{\rm pol})
\label{eq:2.33}
\end{eqnarray}
where $\vert \rho_{\rm pol})=\widehat \Pi \vert \phi_{\rm eff})$;
the final equality follows from an integration by parts, and use
of the continuity equation. Using the fact that $\widehat U$ is
self-adjoint, we obtain the very simple expression for the absorption
\be{2.34}
\biggl\langle \overline{{dE\over{dt}}}\biggr\rangle
={\textstyle{1\over 2}}\, {\rm Re}\, ({\bf j}|{\bf E})
\ .
\ee
Using the continuity equation and 
\req{2.20}, \req{2.26} and \req{2.27},
\be{2.35}
({\bf j}|{\bf E}) = %e^{-1} 
{\rm i}\omega\,
(\phi_{\rm eff}|\widehat \Pi^+(\omega)|\phi_{\rm eff})
+({\bf E}_{\rm ind}|\widehat \sigt^+(\omega)|{\bf E}_{\rm ind})
\ .
\ee
This gives our final expression for the absorption
\be{2.36}
\biggl\langle \overline{{dE\over{dt}}}\biggr\rangle
={\textstyle{1\over 2}}\,\omega \,% e^{-1} 
{\rm Im}\,
(\phi_{\rm eff}|\widehat \Pi(\omega)|\phi_{\rm eff})
+ {\textstyle{1\over 2}}\, {\rm Re}\,
({\bf E}_{\rm ind}|\widehat \sigt(\omega)|{\bf E}_{\rm ind})
\ .
\ee
These are two independent contributions to the
the rate of absorption, depending on the electric
and magnetic fields respectively. It is not obvious that
these are correctly identified as the electric and magnetic dipole
coefficients, because the electric field may induce a charge
density with non-zero angular momentum. We will now show that
the first term is due solely to the electric dipole.
Using the continuity equation and an integration by parts, we 
find
\be{2.37}
{\rm Re}({\bf j}\vert {\bf E}_{\rm ext})=
\omega\, {\rm Im}(\rho \vert \phi_{\rm ext})=
\omega\, {\rm Im}\bigl[{\bf d}.{\bf E}_{\rm ext}\bigr]
\ .
\ee
The electrically induced absorption therefore depends only upon
the induced dipole moment, and is independent of the magnetic
moment induced by the electric field.

\section{Equivalence with perturbation theory}

\subsection{An alternative
expression for the the absorption coefficient}

In this section, we concentrate
on the electric absorption. We describe an alternative
approach to calculating the absorption coefficient, which
was used in [4-7], and show that it is equivalent to
the first term in \req{2.37} provided the polarisation
operator $\widehat \Pi(\omega)$ is related in a simple
way to a propagator $\widehat P(\omega)$. This relation
will be established in section III B.

We will consider the action of the 
effective potential $\phi_{\rm eff}({\bf r},t)$ on the electrons.
We may use either quantum mechanical or classical perturbation
theory. We will describe the quantum mechanical approach, and will 
use semiclassical approximations: a classical theory in which
quantum mechanics only enters in choosing the Fermi-Dirac
distribution for the initial distribution of electrons gives
identical results. 
Conceptually, the simplest method for calculating the absorption 
in the first using the Fermi golden rule. This is expressed
in terms of matrix elements $\phi_{nm}$ of the perturbation 
in the basis $|\psi_n\rangle$
formed by the eigenstates of the single-particle
effective Hamiltonian, $\hat H_{\rm eff}$:
\begin{eqnarray}
&&\hat H_{\rm eff}\ket {\psi_n}=E_n\ket{\psi_n}\nonumber\\
&&\phi_{nm}=\bra {\psi_n}\hat \phi \ket {\psi_m},\ \ \ 
\hat \phi =\phi_{\rm eff}(\hat {\bf r})
\label{eq:3.1}
\end{eqnarray}

The Fermi golden rule states that the rate of transition 
from an initially occupied state to a quasi-continuum of 
final states, with density of states $n$ and with energy
differing by $\hbar \omega$ from the original state, is given by 
\be{3.2}
R={\pi n e^2\over{2 \hbar}} \langle \vert \phi_{nm} \vert^2 \rangle
\ee
where the angle brackets denote an average over matrix elements
$\bra{\psi_n}\hat \phi_{\rm eff}\ket{\psi_m}$.
We will consider the case where both the temperature and the 
photon energy are small compared to other energy scales in the 
problem; generalisations are straightforward. Absorption
of energy occurs due to the excitation of electrons in 
occupied states below the Fermi level to empty states
above the Fermi level. The number of states which can be excited
is $\sim n\hbar \omega$, and the energy absorbed in each
transition is $\hbar \omega$: the total rate 
of absorption of energy is given by multiplying these factors 
by the transition rate $R$, giving
\be{3.3}
\overline{dE\over {dt}}={\textstyle {1\over 2}}\pi\hbar n^2 e^2 \omega^2
\, \langle \vert \phi_{nm}\vert^2\rangle
\ .
\ee
Both \req{2.37} and \req{3.3} are quadratic functions of $\phi_{\rm eff}$,
but it is not immediately clear how they can be related. 
We will now discuss why they are equivalent.

The mean-square matrix element can be estimated from the 
correlation function $C_{\phi \phi}(t)$ of the effective 
potential
\be{3.4}
\langle \vert \phi_{nm}\vert^2 \rangle 
={1\over{\pi \hbar n}}{\rm Re}
\int_0^\infty\!\!dt\,\,
{\rm e}^{{\rm i}\omega t}\,
C_{\phi \phi}(t)
\ee
where the correlation function is defined by
$$C_{AB}(t)=
\langle A({\bf r},{\bf p})B({\bf r}_t,{\bf p}_t)\rangle
\equiv {1\over{\Omega'(E)}}\int\!d{\bf r}\int\!d{\bf p}\ 
A({\bf r},{\bf p})B({\bf r}_t,{\bf p}_t)
\delta [E_{\rm F}-H({\bf r},{\bf p})]$$
\be{3.5}
\Omega'(E)=\int\!d{\bf r}
           \int\!d{\bf p}\ \delta[E-H({\bf r},{\bf p})]
\ee
and ${\bf r}_t$, ${\bf p}_t$ are the phase space coordinates
evolved under the Hamiltonian dynamics for time $t$, starting
from the point $({\bf r},{\bf p})$.
It will be convenient to define a propagator 
$P({\bf r},{\bf r}';t)$ which gives the probability of 
reaching ${\bf r}'$ from ${\bf r}$ in time $t$:
\be{3.6}
P({\bf r},{\bf r}';t)
=\langle \delta ({\bf r}_t-{\bf r}')\ \rangle \theta (t)
\ee
where $\theta (t)$ is a step function which makes the propagator
zero for negative time. The averaging will be defined in the next 
subsection.
With this definition we have
\be{3.7}
C_{\phi \phi}(t)={1\over V}\int\!d{\bf r}\int\!d{\bf r}'\ 
P({\bf r},{\bf r}';t)\,\phi ({\bf r})\,\phi({\bf r}')
\equiv(\phi \vert \widehat P(t)\vert \phi )
\ee
where the operator $\widehat P(t)$ is defined by analogy
with \req{2.21}. Introducing the Fourier transform 
$\widehat P(\omega)$
of the propagator, we have
\begin{eqnarray}
\biggl\langle \overline{{dE\over{dt}}} \biggr\rangle
&=&{\textstyle{1\over 2}} \nu e^2 \omega^2\ {\rm Re}
\int_0^\infty \!\!dt\ {\rm e}^{{\rm i}\omega t}
(\phi_{\rm eff}\vert \widehat P(t)\vert \phi_{\rm eff})
\nonumber\\
&=&{\textstyle{1\over 2}} \nu e^2 \omega^2 \ {\rm Re}
(\phi_{\rm eff}\vert \widehat P(\omega)\vert \phi_{\rm eff})
\label{eq:3.8}
\end{eqnarray}
In the next section it will be shown that there is a
general relation between the propagator and the
polarisability operator:
\be{3.9}
\widehat \Pi (\omega)=e\nu [\widehat I+{\rm i}\omega \widehat P(\omega)]
\ee
where $\nu $ is the density of states per unit volume. 
A relation of this form has been given by
Kirzhnitz [13]. We present a detailed
derivation, based on Liouville's equation, below.
If the 
potential $V_{\rm eff}$ appearing in \req{2.14} is constant within
the conducting particle, we may write $n=\nu V$, where $V$ is
the volume of the particle.
Substituting this into \req{2.37} reproduces \req{3.8}, thus 
establishing its equivalence to \req{3.3}.

\subsection{General relation between polarisability and the propagator}

We will now relate the polarisability operator 
$\Pi ({\bf r},{\bf r}';\omega)$ to the probability
propagator $P({\bf r},{\bf r}';t)$, which is the 
probability that an electron, released at ${\bf r}'$ 
with energy equal to the Fermi energy $E_{\rm F}$, will be 
at position ${\bf r}$ after time $t$. The discussion
will be classical; a quantum mechanical derivation
proceeds along similar lines.

Let the phase-space distribution be $f({\bf r},{\bf p};t)$: 
this will, when convenient, be written $f(\alpha,t)$ where
$\alpha =({\bf r},{\bf p})$. The Hamiltonian will be assumed 
to be of the form 
\be{3.10}
H(\alpha,t)=H_0(\alpha)+X(t)H_1(\alpha)
\ee
where we will be interested in the case where 
$H_0={\bf p}^2/2m+V({\bf r})$ and $H_1=\phi ({\bf r})$. 
The perturbation parameter $X(t)$ is assumed to be small, 
so that $f(\alpha,t)$ may be expanded as a series in $X(t)$:
we will be interested in the expansion as far as the first 
order term:
\be{3.11}
f(\alpha,t)=f_0(\alpha)+\int_{-\infty}^t dt'\ 
X(t')\, g(\alpha,t,t')+O(X^2)
\ .
\ee
Substituting into the Liouville equation, 
$\partial_t f=\{f,H\}$, it is found that $f_0$ is a function
of the unperturbed Hamiltonian $H_0(\alpha)$, and that
the kernel $g(\alpha,t,t')$ of the first order term satisfies
\be{3.12}
X(t)\bigl[g(\alpha,t,t')-\{H_1,f_0\}_\alpha\bigr]
+\int_{-\infty}^t dt'\ X(t')
\bigl[\partial_t g-\{H_0,g\}\bigr]_{\alpha,t,t'}=0
\ee
which is valid for all $X(t)$. The first term implies that
\be{3.13}
g(\alpha,t,t)=g(\alpha)=\{H_1,f_0\}_\alpha =\{H_1,H_0\}_\alpha 
{\partial f_0\over{\partial E}}(H_0(\alpha))
\ .
\ee
The second term implies that $dg/dt=0$, where $d/dt$ is the total 
time derivative along a trajectory, so that 
\be{3.14}
g(\alpha,t,t')=g(\alpha,t-t')
={\partial f_0\over{\partial E}}(H_0(\alpha))
{dH_1\over{dt}}(\alpha_{t-t'}(\alpha))
\ .
\ee
The required approximation is then
\be{3.15}
f(\alpha,t)=f_0(H_0(\alpha))
+{\partial f_0\over{\partial E}}(H_0(\alpha))
\int_{-\infty}^t dt'\ X(t'){dH_1\over {dt}}(\alpha_{t-t'})
\ .
\ee
We will use an alternative form, obtained by integration
by parts
\begin{eqnarray}
f(\alpha,t)&=&f_0(H_0(\alpha))+
X(t){\partial f_0\over{\partial E}}(H_0(\alpha)) H_1(\alpha )
\nonumber\\
&-&{\partial f_0\over{\partial E}}(H_0(\alpha))
\int_{-\infty}^t dt'\ \dot X(t')\, H_1(\alpha_{t-t'}(\alpha))\,.
\label{eq:3.16}
\end{eqnarray}
We will assume that the integral converges. For ergodic systems
this requires that the microcanonical average of $H_1(\alpha )$
vanishes. 
The density of available states in phase space is $(2\pi \hbar)^{-d}$,
where $d$ is the number of degrees of freedom. 
For a system of 
fermions, the appropriate density function is 
$f_0(\alpha,X)=\theta [E_{\rm F}-H(\alpha,X)]/(2\pi \hbar)^d$, 
where $\theta (x)$ is the Fermi-Dirac distribution,
which can be approximated by a downward step function
when the temperature is small compared to the Fermi 
temperature.

Now the charge density of electrons is
\be{3.17}
\rho({\bf r},t)=e\int\!d{\bf p}\ f({\bf r},{\bf p};t)\,.
\ee
The number density of electrons $N(E_{\rm F},{\bf r})$ and the density
of states per unit volume at the Fermi surface $\nu(E_{\rm F},{\bf r})$ are
\be{3.18}
N(E_{\rm F},{\bf r})={1\over{(2\pi \hbar)^d}}
\int d{\bf p}\ \theta (H_0({\bf r},{\bf p})-E_{\rm F})\ , \ \ \ 
\nu(E_{\rm F},{\bf r})=\frac{\partial}{\partial {E_{\rm F}}}N(E_{\rm
F},{\bf r})
\ee
respectively. Also, the local average of any quantity 
$A({\bf r},{\bf p})$ for electrons at the Fermi surface 
is defined as
\begin{eqnarray}
\langle A\rangle_{E_{\rm F},{\bf r}}
&=&\int d{\bf p}\ A({\bf r},{\bf p})\delta (H_0({\bf r},{\bf p})-E_{\rm F})
\bigg/\int d{\bf p}\ \delta (H_0({\bf r},{\bf p})-E_{\rm F})
\nonumber\\
&=&{1\over{(2\pi \hbar)^d \nu(E_{\rm F},{\bf r})}}
\int d{\bf p}\ A({\bf r},{\bf p})\delta (H_0({\bf r},{\bf p})-E_{\rm F})
\label{eq:3.19}
\end{eqnarray}
>From \req{3.16} and the definition \req{3.17}, we have:
\begin{eqnarray}
\rho({\bf r},t)&\sim& e N(E_{\rm F},{\bf r})+eX(t)
\nu(E_{\rm F},{\bf r})\phi({\bf r})
\nonumber\\
&-&e\,\nu(E_{\rm F},{\bf r})\int_{-\infty}^t\!\! dt'\ \dot X(t')
\int\!d{\bf r}'\int\!d{\bf p}\ \delta[E-H_0({\bf r},{\bf p})]\,
\delta[{\bf r}'-{\bf r}_{t-t'}({\bf r},{\bf p})]\,\phi({\bf r}')
 +O(X^2)
\nonumber\\
&=&eN(E_{\rm F},{\bf r})+eX(t) \nu(E_{\rm F},{\bf r})\phi({\bf r})
\nonumber\\
&-&e\nu(E_{\rm F},{\bf r})\int_{-\infty}^t dt'\ \dot X(t')
\int d{\bf r}'\ 
\langle\delta \bigl[ {\bf r}_{t-t'}({\bf r},{\bf p})-{\bf r}'\bigr] 
\rangle_{E_{\rm F},{\bf r}}\ \phi ({\bf r}_{t'})+O(X^2)
\label{eq:3.20}
\end{eqnarray}
With the definition of the propagator
\be{3.21}
P({\bf r},{\bf r}';t)=\theta(t)\,
\langle \delta[{\bf r}'-{\bf r}_t({\bf r},{\bf p})]\rangle_{E_{\rm F},{\bf r}}
\ee
and recalling the definition of the polarisation operator, \req{2.23}, 
we find:
\be{3.22}
\Pi({\bf r},{\bf r}',t-t')=e\,
\theta (t-t')\nu(E_{\rm F},{\bf r})\biggl[\delta ({\bf r}-{\bf r}')
\delta (t-t')+\partial_t P({\bf r},{\bf r}';t-t')\biggr]
\ee
or alternatively, in the frequency domain
\be{3.23}
\Pi({\bf r},{\bf r}',\omega)
=e\,\nu(E_{\rm F},{\bf r})\bigl[\delta ({\bf r}-{\bf r}')+{\rm i}\omega
P({\bf r},{\bf r}';\omega)\bigr]\ .
\ee
We will introduce an operator $\hat \nu$, which is diagonal
in the position representation, so that 
$({\bf r}\vert\ \hat \nu \vert \phi)=\nu(E_{\rm F},{\bf r})\phi({\bf r})$.
Equation \req{3.23} may then be written in the form
\be{3.24}
\widehat \Pi (\omega)
=e\,\widehat\nu\,\bigl[\widehat I+{\rm i}\omega \widehat P(\omega)\bigr]
\ .
\ee
There is also a relationship between the 
non-local conductivity $\widehat \sigt(\omega)$ and the 
propagator $\hat P(\omega)$,
which has previously been obtained by Serota and co-workers
[10,11] (with an alternative derivation given in [8]). Their 
derivation was specific to the case of
diffusive electron motion, whereas that given above 
also includes the ballistic case.

\section{Particular forms for the polarisability}

\subsection{Spatially homogeneous, ballistic system}

For a spatially homogeneous system, 
$\Pi ({\bf r},{\bf r}';\omega)$ is a function of 
${\bf r}-{\bf r}'$, and is conveniently represented
by its Fourier transform, $\Pi ({\bf q},\omega)$:
in $d$ dimensions
\be{4.1}
\hat \Pi (\omega)={V\over {(2\pi)^d}}
\int d{\bf q}\ \vert \chi_{\bf q})\,
\Pi ({\bf q},\omega)\, (\chi_{\bf q}\vert\ ,\ \ \ 
({\bf r}\vert \chi_{\bf q})
={1\over{\sqrt{V}}}{\rm e}^{{\rm i}{\bf q}.{\bf r}}
\ee
where $V$ is the volume of the system.
In the case where the electron motion is ballistic, the
propagator is, for $d=3$,
\begin{eqnarray}
P({\bf r},{\bf r}';t)&=&{1\over{4\pi R^2}}
\delta (R-v_{\rm F}t)\ ,\ \ \ R=|{\bf r}-{\bf r}'|\,,
\nonumber\\
P({\bf r},{\bf r}';\omega)&=&{1\over{4\pi v_{\rm F} R^2}}\,
{\rm e}^{{\rm i}\omega R/v_{\rm F}}
\label{eq:4.2}
\end{eqnarray}
and the Fourier transform representation of the 
polarisability is 
\be{4.3}
\Pi (q,\omega)=\nu e \left(1-{1\over {2\lambda}}\log 
\bigg \vert {\lambda +1\over{\lambda -1}}\bigg \vert
+ {\rm i} \theta(\lambda-1)\,\frac{\pi}{2\lambda}\right)
\ ,\ \ \ 
\lambda ={qv_{\rm F}\over \omega}
\ ,
\ee
which is the semiclassical limit of Eqs. (12.48a,b) in [9].
In two dimensions, $\Pi (q,\omega )$ is given by 
\begin{equation}
\Pi(q,\omega) = 
e\nu 
\left\{
\begin{array}{ll}
1+{\rm i} (\lambda-1)^{-1/2} & \mbox{for $\lambda > 1$}\\[0.4cm]
1-(1-\lambda^2)^{-1/2} & \mbox{for $\lambda < 1$}\,.
\end{array}
\right .
\end{equation}

\subsection{Low and high frequency limits}

In the low frequency limit, it is immediately clear
from \req{3.24} that the induced charge density is 
$\rho ({\bf r})=\nu e [\phi({\bf r})-\langle \phi \rangle]$
where $\langle \phi \rangle$ is the space average
of $\phi $ over the particle. We shall be only be concerned 
with cases where $\langle \phi \rangle$ vanishes, so that 
we may write 
\be{4.4}
\Pi ({\bf r},{\bf r}';\omega) \sim \nu\,e\, \delta ({\bf r}-{\bf r}')
\ ,\ \ \ \omega \ll \omega_{\rm c}\,.
\ee
For sufficiently high frequencies, and sufficiently far
from the boundary of the particle, the conductivity is
local, with value $\sigma (\omega)$:
\be{4.5}
\Sigma_{ij}({\bf r},{\bf r}';\omega)
=\delta_{ij}\,\delta({\bf r}-{\bf r}')\,\sigma (\omega)
\ee
and the bulk conductivity $\sigma (\omega)$ may, in the case
of diffusive electron motion, be approximated by the Drude 
formula
\be{4.6}
\sigma(\omega)={\nu e^2D\over{1+{\rm i}\omega \tau}}\ , \ \ \ 
\tau ={m\over N}{\partial N\over {\partial E}}D
\ee
where $D$ is the diffusion constant, and $m$ the electron 
effective mass. In the case of ballistic electron motion, 
the bulk conductivity is determined purely by the inertia
of the electrons, and is non-dissipative:
\be{4.7}
\sigma(\omega)={Ne^2\over{{\rm i}m\omega}}
\ .
\ee
When the non-local conductivity can be approximated by
\req{4.5}, the non-local polarisability takes the simple
form
\begin{eqnarray}
\Pi ({\bf r},{\bf r}';\omega)&=&{{\rm i}\sigma(\omega)\over{\omega}}
\nabla_i\nabla_i'\delta ({\bf r}-{\bf r}')
\nonumber\\
&=&-{{\rm i}\sigma(\omega)\over{\omega}}
\nabla^2\delta ({\bf r}-{\bf r}')
\ .
\label{eq:4.8}
\end{eqnarray}
This approximation is expected to be valid when $\omega\gg \omega_{\rm c}$,
and when both ${\bf r}$ and ${\bf r}'$ are much greater than 
a distance $\Lambda $ from the boundary: in the ballistic case
$\Lambda =v_{\rm F}/\omega$, and in the diffusive case 
$\Lambda=\sqrt {D/\omega}$. The same conclusion can also be 
reached by considering the expressions \req{4.3}, \req{4.4} in 
the limit $\lambda \to 0$: for $d=3$ we find that 
$\Pi (q,\omega)\sim {1\over 3}\nu e\lambda^2=\nu ev_{\rm F}^2q^2/3\omega^2$,
which is equivalent to the Fourier transform of \req{4.8} when
the conductivity is given by \req{4.7}.

\subsection{Polarisability close to a boundary}

Next we consider the polarisation charge close to the boundary
of the particle. Here we are concerned with the high frequency
case, $\omega \gg \omega_{\rm c}$. In the low frequency case \req{4.4}
gives an adequate approximation, but our discussion of the 
high frequency case assumed that the conductivity could be
approximated as that of a homogeneous system. Another reason
for discussing the boundary separately is that we expect
that the polarisation charge density may have a singularity
there.

We may assume that for $\omega \gg \omega_{\rm c}$ the polarisability
operator is short ranged. A smooth boundary may therefore 
be approximated locally by a flat surface, $z=0$ in some
local Cartesian coordinates. The polarisability
is given by \req{3.14}, and we approximate the propagator from
${\bf r}'=(x',y',z')$ to ${\bf r}=(x,y,z)$ by the sum of a 
direct contribution and a contribution 
originating from an  image source at 
${\bf r}_{\rm im}'=(x',y',-z')$, so that
\be{4.9}
\rho ({\bf r})=\nu e\biggl[ \phi ({\bf r})+{\rm i}\omega 
\int d{\bf r}'\ \bigl[P({\bf r},{\bf r}';\omega)
+P({\bf r},{\bf r}_{\rm im}';\omega)\bigr]\phi ({\bf r}')\biggr]
\ .
\ee
The charge charge density is concentrated in a narrow 
layer at the surface, and may typically be approximated by
writing
\be{4.10}
\rho ({\bf r})=\rho_s(z) K({\bf S})
\ee
where ${\bf S}$ labels points on the boundary, and $z$ is a 
coordinate normal to the boundary. In this case, the potential
in the neighbourhood of the boundary is of the form
$\phi ({\bf r})=\phi_s(z)K({\bf S})$, where $\phi_s(z)$
satisfies
\be{4.11}
\rho_s(z)=\nu e\biggl[\phi_s(z)+{1\over {\Lambda}}
\int_0^\infty  dz'\ \bigl[
G\bigl((z-z')/\Lambda)\bigr)+G\bigl((z+z')/\Lambda\bigr)\bigr]
\phi_s(z')\biggr]
\ee
where $\Lambda =v_{\rm F}/\omega$, and the function $G(x)$
is easily related to the Fourier transform of 
$\Pi (q,\omega)$.

\subsection{Diffusive electron motion}

In the diffusive case, it is possible to write a useful
eigenfunction expansion for the linear response functions:
for $t>0$ the propagator $P({\bf r},{\bf r}';t)$ satisfies the diffusion
equation $\partial_t P=D\nabla^2 P$, or
$[{\rm i}\omega -D\nabla^2] P({\bf r},{\bf r}';\omega)
=-\delta({\bf r}-{\bf r}')$,
and satisfies the Neumann boundary condition. It can be expressed
in terms of a set of eigenfunctions $\chi_n ({\bf r})$ of the 
Helmholtz equation $(\nabla^2+k_n^2)\chi_n({\bf r})=0$, satisfying
the same boundary condition: $\hat {\bf n}.\nabla \chi_n=0$, where
$\hat {\bf n}$ is a normal vector on the boundary of the particle.
The propagator can then be written
\be{4.12}
\widehat P(\omega)=\sum_n {1\over {{\rm i}\omega-Dk_n^2}}
\vert \chi_n)(\chi_n\vert
\ .
\ee
Expansions for other linear response functions are easily 
obtained in the same form. For example, if the density
of states per unit volume $\nu$ is independent of 
${\bf r}$, \req{3.15} implies that
the polarisability can be written in this form, with the
coefficient of the operator $\vert \chi_n)(\chi_n\vert$ 
given by $\nu eDk_n^2/(Dk_n^2-{\rm i}\omega)$.

\section{The self-consistent field}

\subsection{Approximate equations for the self-consistent fields}

Here we discuss how the solution of the self-consistent equations
can be greatly simplified by the use of \lq semiclassical' approximations.
We consider the electric dipole absorption first. 

Calculation of the electric dipole absorption coefficient via either 
\req{2.37} or \req{3.8} requires the self-consistent fields 
$\phi_{\rm eff}({\bf r},\omega)$, which is given by equation \req{2.30}:
\be{5.1}\vert \phi_{\rm ext})=\bigl[\hat I-\hat U\hat \Pi(\omega)\bigr]
\vert \phi_{\rm eff})
\ .
\ee
Formally, solution of this equation requires calculation
of the inverse of $\hat I-\hat U\hat \Pi(\omega)$: this could 
be done explicitly in a numerical calculation by expanding
in a suitable basis set. We will aim instead for an
approximate analytic solution. For frequencies small 
compared to the plasma frequency $\omega_{\rm p}$, the external 
electric field is \lq screened' by polarisation charges, so 
that the internal field is much smaller than the externally 
applied field. The key physical intuition is that the 
external electric field is almost exactly cancelled by 
the electric field due to the induced charge density
$\rho({\bf r})$. Let $\rho_{\rm cl}({\bf r})$ be the
charge density induced on the particle by a static 
external field, according to classical electrodynamics:
this charge density gives an induced electric field
which precisely cancels the externally applied field
inside the particle. For frequencies small compared to
the plasma frequency, the induced charge density
is well approximated by $\rho_{\rm cl}({\bf r})$:
we will assume that
\be{5.2}
\vert \rho)=\vert \rho_{\rm cl})+O(\omega/\omega_{\rm p})+O(a/a_0)
\ee
where $a$ is the characteristic dimension of the particle,
and $a_0$ is the Bohr radius. The classical charge
distribution formally satisfies an equation analogous
to \req{5.1}, in which the term representing the internal 
field $\vert \phi_{\rm eff})$ is set equal to zero:
\be{5.3}
\vert \phi_{\rm ext})+\hat U\vert \rho_{\rm cl})=0
\ .
\ee
We will denote our approximation to the effective potential 
$\vert \phi_{\rm eff})$ by $\vert \phi)$: it is the potential
which generates the polarisation charge $\vert \rho_{\rm cl})$,
and is given by
\be{5.4}
\vert \rho_{\rm cl})=\hat \Pi (\omega)\vert \phi)
\ee
or equivalently by 
$\vert \phi_{\rm ext})+\hat U\hat \Pi(\omega)\vert \phi )=0$.
Comparing with \req{5.1}, it is clear that this solution
$\vert \phi )$ is a good approximation to $\vert \phi_{\rm eff})$
provided $\vert\vert \hat U\hat \Pi (\omega)\vert\vert \gg 1$, where
$\vert\vert \hat X\vert\vert$ is an appropriate norm of the 
operator $\hat X$. To estimate this norm, we consider the effect
of an arbitrary potential $\phi $: at zero frequency, the 
induced charge density is $\hat \Pi\phi =e\nu \phi$, and for
a particle of characteristic dimension $a$ in $d$ dimensions,
the induced charge may be approximated
by a dipole formed by charges of magnitude $Q\sim \rho a^d$,
with separation $a$: this results in an electrical potential
of magnitude $\phi'\sim eQ/(\epsilon_0 a)$. In three 
dimensions this leads to the following estimate for
$\vert\vert \hat U\hat \Pi \vert\vert \sim \phi'/\phi$
\be{5.5}
\vert\vert \hat U\hat \Pi \vert\vert \sim 
{\omega_{\rm p}^2\over{\omega^2}}
\ee
where $\omega_{\rm p}$ is the three dimensional bulk plasma frequency,
$\omega_{\rm p}=[Ne^2/(4\pi \epsilon_0 m)]^{1/2}$, $N$ being the 
electron density. For frequencies 
$\omega\gg \omega_{\rm c}$, a similar argument gives
\be{5.6}
\vert\vert \hat U\hat \Pi (\omega\gg \omega_{\rm c})\vert\vert
\sim {\omega_{\rm p}^2\over{\omega \omega_{\rm c}}}
\ .
\ee
In the case of ballistic dynamics $\omega_{\rm c} = v_{\rm F}/a$, 
equations \req{5.5} and \req{5.6} are of the order
of $\sigma_0/(\epsilon_0\omega_{\rm c})$ and  $\sigma_0/(\epsilon_0\omega)$
respectively.

We can, in principle, determine improved approximations
to the exact solution of \req{5.2} from the solution of \req{5.4}:
\begin{eqnarray}
\vert \phi_{\rm eff})&=&
-\bigl[\hat I-\hat U\hat \Pi(\omega)\bigr]^{-1}
\hat U\hat \Pi(\omega)\,\vert \phi)
\nonumber\\
&=&\vert \phi)
-\bigl[\hat I-\hat U\hat \Pi(\omega)\bigr]^{-1}\,\vert \phi)
\ .
\label{eq:5.7}
\end{eqnarray}
Equations \req{5.5} and \req{5.6} show that the correction term
in \req{5.7} is small. Having shown that 
$\vert \phi_{\rm eff})\sim \vert \phi)$, we discuss how to
estimate the solution $\vert \phi )$ of \req{5.4}
in Secs. VB and VC.

Finally we consider the semiclassical solution for the 
magnetically induced field, ${\bf E}_{\rm ind}$ which
satisfies \req{2.29}:
\be{5.8} 
\nabla .[\hat I-{\rm i}\omega \epsilon_0\widehat \sigt(\omega)]\,
\vert {\bf E}_{\rm ind})=0
\ .
\ee
At low frequencies we can estimate the conductivity by 
$\Sigma \sim Ne^2/m\omega_{\rm s}$ where $\omega_{\rm s}$ is the 
scattering frequency; at high frequencies $\omega $
is replaced by the frequency $\omega $. In the low frequency
limit we therefore estimate
\be{5.9}
{1\over {\epsilon_0 \omega}}\vert \vert \widehat\sigt (\omega) \vert \vert 
\sim {\omega_{\rm p}^2\over {\omega \omega_{\rm s}}}
\ .
\ee
In the frequency range that we are concerned with, the term 
involving the identity operator in \req{5.8} is therefore negligible;
the same conclusion applies when $\omega \gg \omega_{\rm s}$. We can 
therefore find an approximate solution to \req{5.8} by requiring
that the induced charge density is zero, i.e.
\be{5.10}
\nabla .\widehat \sigt  (\omega) \vert {\bf E}_{\rm ind})=0
\ .
\ee
This justifies the neglect of a cross-term in \req{2.1} with
an electric dipole induced by the magnetic field.
A solution to \req{5.10} may be determined by choosing an
initial approximation ${\bf E}_{\rm ind}'$ which
satisfies 
$\nabla \wedge {\bf E}_{\rm ind}'={\rm i}\omega {\bf B}_{\rm ext}$.
A polarisation charge $\rho'_{\rm ind}$ would be generated 
from this field. An additional field which is the gradient
of a potential $\chi ({\bf r})$ is added, such that
${\bf E}_{\rm ind}={\bf E}_{\rm ind}'+\nabla \chi$. The 
condition upon $\chi $ for \req{5.10} to be satisfied is
\be{5.11}
\vert \rho_{\rm ind}')=\widehat \Pi (\omega)\, \vert \chi )
\ .
\ee
This equation is analogous to \req{5.4}.

\subsection{Solutions in the ballistic case} 

We will discuss approximate solutions of \req{5.4} valid in
the limits $\omega \ll \omega_{\rm c}$ and $\omega \gg \omega_{\rm c}$.
The first of these represents the static potential required
to hold the classical charge distribution in place in the 
zero-frequency limit: it will be written $\vert \phi_{\rm stat})$,
and its form is immediately apparent from \req{3.24}:
\be{5.12}
\vert \phi_{\rm stat})={1\over{e\nu}}\vert \rho_{\rm cl}) \ .
\ee
This is simply a linearised Thomas-Fermi approximation [12].
The semiclassical approximations underlying this expression
assume that the potentials are slowly varying on the scale of 
the Fermi wavelength. This is questionable when the charge density
has a singularity on the surface of the particle: we will 
return to this later.

In the limit $\omega \gg \omega_{\rm c}$, we found [equation \req{4.8}] 
that the polarisability may be approximated by 
$\hat \Pi \sim -({\rm i}\sigma (\omega)/\omega)
\nabla^2_{\bf r}\delta({\bf r}-{\bf r}')$, for 
${\bf r}$ and ${\bf r}'$ not too close to the boundary.
For points not too close to the boundary, or where 
the charge density is non-analytic, we can approximate 
the solution of \req{5.4} by a \lq dynamic' potential,
which is of the form 
$\vert \phi_{\rm dyn})=\lambda \vert \psi)$,
where $\nabla^2 \psi ({\bf r})=\rho_{\rm cl}({\bf r})$,
and $\hat n.\nabla \psi({\bf r})=0$. Substituting
these forms into \req{5.4} we find that 
$\lambda=-{\rm i}\omega/\sigma(\omega)$:
\be{5.13}
\vert \phi_{\rm dyn})=-{{\rm i}\omega\over{\sigma(\omega)}}
\vert \psi)\ , \ \ \ \nabla^2\vert \psi)=\vert \rho_{\rm cl})
\ .
\ee
An interpretation of the dynamic potential is that it moves 
the polarisation charge into place.

Close to the boundary \req{5.13} is not necessarily a good
approximation to the effective potential. One reason is
that the approximations underlying \req{4.8} fail, and 
the polarisation must be described by \req{4.9} or \req{4.11}.
Another reason is that
the charge density $\vert \rho_{\rm cl})$ has a singularity
there. In the notation of \req{4.10}, the projected charge 
density $\rho_s(z)$ is concentrated on the boundary in 
the three dimensional
case [4] so that $\rho_s(z)\sim \delta(z)$, and in the two dimensional 
case it diverges on the boundary, such that 
$\rho_s(z)\sim z^{-1/2}$ for $z>0$ [6]. The form of \req{4.11}
indicates that the potential $\phi_s(z)$ also has the same
type of singularity as the charge density at the boundary.

To summarise, the following picture emerges. For low frequencies,
$\omega \ll \omega_{\rm c}$ the potential is approximately
$\phi_{\rm stat}({\bf r})$. At high frequencies, 
$\omega \gg \omega_{\rm c}$, the potential is well approximated 
by $\phi_{\rm dyn}({\bf r})$ within the interior of the 
particle. In the vicinity of the boundary, the potential 
has a dominant divergent contribution, which is well approximated by
$\phi_{\rm stat}({\bf r})$.

\subsection{Diffusive case}

In the diffusive case, equation \req{5.4} can be solved 
exactly, using the representation of the propagator
in the form \req{4.12}. Expanding the potential $\vert \phi)$ in 
terms of the functions $\vert \chi_n)$ leads to the expression
\be{5.14}
\vert \phi )={1\over {e\nu}}\sum_n 
(\chi_n \vert \rho_{\rm cl})
\biggl[1-{{\rm i}\omega\over{Dk_n^2}}\biggr]  
\vert \chi_n)
=\vert \phi_{\rm stat})+\vert \phi_{\rm dyn})
\ee
Note that in the diffusive case the potential is precisely
equal to the sum of the  static and  dynamic contributions
[7].

\section{Discussion: calculation of the absorption coefficient}

Once an adequate approximation for the effective potential 
has been obtained, the electric absorption coefficient is obtained 
from \req{3.8}: the absorption coefficient is proportional
to $\omega^2(\phi\vert \hat P(\omega)\vert \phi )$.
Previous papers [4-8] have discussed methods for the evaluation of the 
absorption coefficient using equation \req{3.8} in some detail, for 
specific cases. This paper has presented a general approach
to the determining of the effective potential, and some 
remarks on applying this to calculating the absorption coefficient 
may be useful.

When the electron motion is diffusive, the absorption 
is very easily evaluated using \req{5.14} and
\req{4.12}. It is found that the coefficient
is proportional to $\omega^2$, and that (at least within 
the framework of the approximations used in section V) the frequency
scale $\omega_{\rm c}$ plays no role. The absorption coefficient
can be shown to be exactly equal to the classical value in this
case [7,5]. 

The case of ballistic electron motion is more difficult.
It might be expected that 
$C_{\phi\phi}(\omega)=(\phi \vert \hat P(\omega)\vert \phi)$ 
approaches a non-zero limit as $\omega \to 0$, implying that 
the absorption coefficient is proportional to $\omega^2$ for 
low frequencies. This expectation is correct for cases
where the electron motion is ergodic (the most important cases 
being diffusive electron motion,
and the ballistic case with a rough boundary). In the case
of integrable electron motion, which can be realised experimentally
if the boundary appears smooth on the scale of the Fermi wavelength,
$C_{\phi \phi}(\omega)$ typically approaches zero as $\omega \to 0$
in a non-analytic fashion. In the important special case
of particles with circular symmetry, $C_{\phi\phi}(\omega)$
is zero for $\omega <\omega_{\rm c}$, where $\omega_{\rm c}$ is the frequency
of a glancing circular orbit. Thus the electron dynamics
plays an important role in determining the low frequency
absorption.

The high frequency absorption, by contrast, is determined by the 
nature of the singularities of the function $f(t)=\phi({\bf r_t})$,
which can result from singularities in the motion ${\bf r}(t)$,
or in the potential $\phi({\bf r})$. We will discuss the ballistic
case. The dominant contribution comes from the singularities 
of $\phi({\bf r})$ in the neighbourhood of the boundary. 
The absorption coefficient was calculated in [4] and [6] 
for the three dimensional and two dimensional cases respectively, 
assuming that the effective potential is equal to 
$\phi_{\rm stat}({\bf r})$. In three dimensions, 
this potential is a delta function singularity
concentrated on the boundary, and in two dimensions it
diverges as $z^{-1/2}$, where $z$ is the normal distance
from the boundary. These forms for the potential imply
that $(\phi \vert \hat P(\omega)\vert \phi)\sim \omega^{0}$
in three dimensions and $\sim \omega^{-1}$ in two dimensions.
In [4] and [6] it was predicted that in the case of particles with 
circular symmetry, the absorption coefficient shows a sequence 
of resonances superposed on a regular contribution increasing as 
$\omega^2$ and $\omega$ for three and two dimensions respectively.

The more sophisticated approach introduced in this paper indicates 
that the potential $\vert \phi )$ should satisfy \req{5.4}, whereas
the potential used in [4], [6] was simply the static potential,
satisfying $\vert \rho_{\rm cl})=\hat \Pi (0)\vert \phi_{\rm stat})$.
We must discuss the extent to which this refinement will change the 
results. In section V B we argued that the singularities
of the effective potential at the boundary are the same
as those of the \lq static' potential.
We can therefore hypothesise that the more refined theory would make a 
quantitative rather than qualitative difference to the results.
This issue will be addressed in a subsequent paper, which
will consider the inversion of \req{4.11} to determine the 
potential $\phi_s(z)$ from the charge density $\rho_s(z)$,
and its use to estimate the high frequency absorption coefficient.
\acknowledgments{
BM would like to acknowledge support of the SFB393 (project C6),
the work of MW was supported by a
research grant from the EPSRC, 
and GR/L02302.}


\begin{thebibliography}{}
\bibitem{carr85}
G. L. Carr, S. Perkowitz, and D. B. Tanner, in {\sl Infrared and
Millimeter Waves}, {\bf 13}, 169, ed. K. J. Button, Academic Press, (1985).
\bibitem{per81}
\bref{2}{J. A. A. J. Perenboom, P. Wyder and F. Meier}{Phys. Rep.}{78}
{173}{1981} 
\bibitem{dev89}
\bref {3}{R. P. Devaty}{Physica}{A157}{262-68}{1989}
\bibitem{aus93}
\bref {4}{E. J. Austin and M. Wilkinson}{J. Phys.: Condens. Matter}
{5}{8461-8484}{1993}
\bibitem{meh97}
\bref {5}{B. Mehlig and M. Wilkinson}{J. Phys.: Condens. Matter}
{9}{3277-90}{1997}
\bibitem{wil94}
\bref {6}{M. Wilkinson and E. J. Austin}{J. Phys.: Condens. Matter}
{6}{4153-4166}{1994}
\bibitem{wil98}
\bref {7}{M. Wilkinson and B. Mehlig}{Eur. J. Phys.}{1B}{397-8}{1998}
\bibitem{wil98b}
\bref {8}{M. Wilkinson, B. Mehlig and P. N. Walker}
{J. Phys.: Condens. Matter}{10}{2739-58}{1998}
\bibitem{fet71}
A. L. Fetter and J. D. Walecka, {\sl Quantum Theory of Many-particle 
Systems}, New York: McGraw Hill, (1971).
\bibitem{kan88}
\bref{10}{C. L. Kane, R. A. Serota and P. A. Lee}{Phys. Rev.}{B37}{6701-10}
{1988}
\bibitem{ser90}
\bref{11}{R. A. Serota, J. Yu and Y. H. Kim}{Phys. Rev.}{B42}{9724-7}{1990}
\bibitem{ash76}
N. W. Ashcroft and N. D. Mermin, {\sl Solid State Physics},
Philadelphia: Saunders College, (1976).
\bibitem{kir75}
D. A. Kirzhnitz, Yu. E. Lozovik, and G. V. Shpatakovskaya,
Sov. Phys. Usp. {\bf 18}, 649-71, (1975)
\end{thebibliography}
\end{document}